\documentclass[showpacs,showkeys,floatfix]{revtex4}
\usepackage[dvips]{graphicx}
\usepackage{epsfig}

\usepackage[cp1251]{inputenc}
\usepackage[english, russian]{babel}
\usepackage{color}
\usepackage{ulem}
\usepackage{bm}

\begin{document}

\title{Topological metastability of textures in biaxial nematics.}

\author{V.L. Golo} 
\affiliation{Department of Mechanics and Mathematics, Lomonosov Moscow
State University\\ Moscow, Russia, and \\
National Research University Higher School of Economics\\
Moscow, Russia }
\email{voislav.golo@gmail.com}

\author{E.I. Kats}
\affiliation{Landau Institute for Theoretical Physics \\
Chernogolvka, Moscow region, Russia}    
\email{efim.i.kats@gmail.com}

\author{D.O. Sinitsyn} 
\affiliation{Semenov Institute of Chemical Physics\\
Moscow, Russia}
\email{d_sinitsyn@mail.ru}

\begin{abstract}
    We consider textures of biaxial nematics confined between two parallel plates.
    The boundary conformations at the bordering plates are supposed to be identical,
    the gradients of the order parameter being generally nonzero.
    We claim that for any texture (including stable uniform order parameter alignment) there exists its
    counterpart texture which is also a minimum of the gradient elastic energy. Our arguments  are   based on the topological analysis of the conformation of the order parameter.
\end{abstract}



%\pagebreak
%\documentclass{article}
%\textwidth=130mm \hfuzz=1.5pt \textheight=175mm
%\usepackage[dvips]{graphicx}
%\usepackage{epsfig}
%\usepackage{longtable}
%\usepackage{float}
%\parskip 3mm
%\sloppy

\maketitle

%\pagebreak

It was de Gennes who suggested the matrix form of the order parameter for liquid crystals, \cite{gennes}.
For uniaxial nematics it reads
$$
  A_{ij} = const \cdot \left( \frac{1}{3} \delta_{ij} \; - \; n_i n_j      \right)
$$
where $n_i, n_j$ are coordinates of the unit vector ${\bf n}$.  For the general case of the biaxial nematic (see e.g., \cite{DM06}, \cite{VP08},
\cite{PD12}) the 
order parameter is a real symmetrical $3 \times 3$ - matrix with the trace zero, characterized by five independent parameters.  
The main point about the matrix
of the order parameter is that it can be cast in the form 
\begin{equation}
 \label{eq:orbit}
 {\hat A} = {\hat R}^{-1} {\hat A}_0 {\hat R}
\end{equation}
where ${\hat R}$  is a rotation matrix, and ${\hat A}_0$ reads
\begin{equation}
 \label{eq:a_0}
    {\hat A}_0 =\left( 
	    \begin{array}{ccc}
	        \lambda_1 &   0          &  0                              \\
		 0         &   \lambda_2 &  0                              \\
		 0         &  0          &  - ( \lambda_1   +  \lambda_2 ) \\
	    \end{array}
        \right)   
\end{equation}
It should be noted that the order parameter in the form given by (\ref{eq:orbit}) have been used
by Monastyrsky and Sasorov, \cite{monst1}, for studying the topology of defects in biaxial liquid crystals
(for a bit different topological approach see earlier review paper \cite{KL88} and many relevant references there).

In the present note we study conformations of the one-dimensional textures in biaxial nematics, 
which can be considered as a realistic model for liquid crystalline films confined
between uniform glass substrates.  To that end we have to consider
minima of the gradient part of the free energy
$$
      {\cal F}_{\nabla} = \int F_{\nabla} dV
$$
The density $F_{\nabla}$ for the one-dimensional case reads
\begin{equation}
      \label{eq:grad}
      F_{\nabla} = K_1 Tr \left( 
                                \partial_z {\hat A} \cdot    \partial_z {\hat A} 
			  \right)
                          +  K_2 \left( 
			                \partial_z {\hat A} \cdot    \partial_z {\hat A}  
				 \right)_{33}
\end{equation}
where $z=x_3$ is the spatial coordinate of the texture, and $\partial_z$ is the derivative with respect to $z$,
and $K_1$, $K_2$ are two phenomenological elastic constants.
The boundary conditions we have assumed at the plates of a cell read ${\hat A}(z=0) = {\hat A}(z=L)$, 
that is one may consider them as parallel.

The values of the order parameter for a liquid crystalline phase determined 
by equations (\ref{eq:orbit})(\ref{eq:a_0}) 
form a submanifold ${\bf \Re}$ in the ambient space ${\bf R}^5$.  It is the standard exercise to see that
in the general case of the eigenvalues of the matrix of the order parameter being unequal, the space 
$\bf \Re$ is in the one-to-one correspondence with the manifold of three-dimensional rotations,
the Lie group $SO(3)$, that is, topologically,  the tree-dimensional real projective space $RP^3$.   
But it should be noted that they may have different riemaniann metrics. Thus,we may infer 
that the fundamental group $\pi_1$ of $\bf \Re$ is not trivial. In fact, it is of the second order
\begin{equation}
      \label{eq:group}
      \pi_1({\Re })  = {\bf Z}_2
\end{equation}
The above equation, or the isomorphism of groups, is of primary importance for what follows.

It is worth noting that the set of real symmetrical $3 \times 3$ matrices with trace zero is 
a real 5-dimensional space, ${\bf R}^5$, and the quadratic form
\begin{equation}
      \label{eq:metric}
            K_1 Tr \left(  X \cdot X  \right) \; + \;  K_2 \left( X \cdot  X  \right)_{33}
\end{equation}
defines an euclidean metric for it, $X$ being a matrix of the kind indicated above. 
The metric generates the riemaniann metric, or the arc length, given by the equation
\begin{equation}
      \label{eq:arc}
       {\cal A} = \int\limits_{z=0}^{z=Z} \; \sqrt{ F_{\nabla} } \, dz
\end{equation}
on the manifold $\Re$. 

The circumstance enables us to  employ the topological theory for finding minima 
of the one-dimensional free energy.
To that end let us note that for one thing the functional of the gradient energy of a one-dimensional texture
\begin{equation}
       \label{eq:lagrange}
       {\cal L} = \int\limits_{z=0}^{z=Z} \; F_{\nabla} \, dz
\end{equation}
may be visualized as the Lagrangian for a free particle on the manifold $\Re$,
and for another it is closely related to that of the arc length (\ref{eq:arc}).
It is easy to see that the Euler-Lagrange equations 
of (\ref{eq:arc}) and (\ref{eq:lagrange}) coincide.   Consequently, their variational equations, which 
describe the second variation of the functionals, also coincide.

It is important that for both functionals the matrices of second derivatives of the Lagrangian with respect
to the time derivatives of the variables are positively defined. In fact, as far as (\ref{eq:lagrange}) is
concerned it is the matrix of the kinetic energy of the particle.  Consequently, according to the Jacoby criterion
a solution to the Euler equations  connecting two points $P_1$ and $P_2$ is the minimal one if it does not contain 
pairs of conjugate points;  two points $Q_1$ nad $Q_2$ being conjugate if there is a 
solution of the variational equations for the Euler equations of a functional, 
equal to zero both at $Q_1$ nad $Q_2$, \cite{gelffom}.

Thus, a solution  is the minimal one if it is minimal  either for equations (\ref{eq:arc}) or  (\ref{eq:lagrange}).

The homotopy theory provides a clew to finding closed geodesics and hence periodic solutions to the equations for textures.
It claims, \cite{bishop}, that if the fundamental group, $\pi_1$, of a manifold is not trivial, 
a homotopy class of closed curves generating an element $x \ne 1 $ of $\pi_1$ contains a closed minimal geodesics.
According to (\ref{eq:group}) this is the case for $\Re$, therefore there should be stable 
textures with periodic boundary conditions, in biaxial nematics.  In fact we should call them metastable for they 
may correspond to minima which are not the lowest ones.  
This is the main result of our present work. We claim that topology predicts
mandatory existence of metastable (i.e. the minima of the gradient energy) textures in biaxial nematics. The result is valid irrespective of explicit
analytical form of the gradient energy, including multi-constant anisotropic elastic energy, when analytic
approach does not work, and even numeric methods are prohibitively time consuming.

It is alleged to be known that metastable textures corresponding to
nontrivial elements of $\pi_1$ do exist in uniaxial nematics,\cite{gennes} 
and have important technological applications. 
The topological arguments presented above can be employed in their case  without any
significant modifications.
It would be interesting to create in real life the biaxial textures suggested above. 
They could provide a wider scope of the optical applications.

V.L. Golo acknowledges the support of the Programme Process of Basic Research, 
National University Higher School of Economics.

E.Kats thanks RFBR project 13-02-00120 supporting his work.

\end{document}